# What drives Oregon shelf summer hypoxia?

A.O. Koch[1], Y.H. Spitz[2], H.P. Batchelder[2,3].

1. University of Southern Mississippi, Department of Marine Science
2. College of Oceanic and Atmospheric Sciences, Oregon State University
3. North Pacific Marine Science Organization, Sidney, BC, Canada

Corresponding author: A. O. Koch, Sequenom Integrated Genetics, LabCorp, 3595 John Hopkins Ct, San Diego, CA 92121, USA. (andreyokoch@gmail.com)

**Abstract.**

Using coupled biological-physical model based on NPZD-type biological model and 3D coastal ocean model (ROMS) we studied dissolved oxygen (DO) dynamics and hypoxia development on Oregon shelf during April-August of 2002, 2006, and 2008. We found that shelf hypoxia existed during summer months of all three years. It was characterized by variable severity, horizontal and vertical extent, duration and timing, and it was more pronounced in 2002 and 2006. By the means of numerical sensitivity analysis we found out that: inadequate initial DO and NO3 conditions in late-spring 2002 prevented or delayed hypoxia development; offshore and especially northern DO and NO3 boundary conditions are important to simulating hypoxia on the Oregon shelf, this was especially critical for early bottom hypoxia on the shelf north of 45$^o$N in 2006; hypoxia occurred earlier in the north in 2006 and in the south (Heceta Bank) in 2002, perhaps, due to different northern boundary conditions for these years; the DO and NO3 conditions at western open boundary located some 400 km offshore are unimportant for DO dynamics in spring-summer. Although DO production due to biological processes is large, physical processes, mostly horizontal advection and diffusion, are responsible for net DO reduction in spring-summer and hypoxia onset in summer on the Oregon shelf. The physical mechanism most responsible for Oregon shelf hypoxia is the coastal upwelling. Diffusive fluxes of NO3 and DO are negligible at northern and southern boundaries of the Oregon shelf and appreciable at the western boundary. In 2006, about two thirds of total April-August DO loss happened in April-May as a result of strong and long-lasting upwelling event.

**Introduction.**

Low dissolved oxygen (DO) conditions or hypoxia in coastal areas of the World Ocean has become an alarming societal and economical issue in the last decades [Scientific Assessment of Hypoxia, 2010]. According to Diaz and Rosenberg [2008] the number of dead zones in the coastal oceans where concentration of nutrients is high and DO concentration is very low have increased exponentially worldwide since 1960s. Hypoxic DO concentrations in deep ocean are attributed to vertical oxygen structure and are generally referenced as ocean minimum zones (OMZ) with thicknesses from 700 m to 3500 m and upper boundary in a depth range of 150 – 850 m [Paulmier and Ruiz-Pino, 2009]. In the North-Eastern Pacific Ocean the upper boundary of OMZ core with DO<= 0.5 ml/l is found at 600-800 m [Paulmier and Ruiz-Pino, 2009]. Although OMZs are not seen directly as hypoxia in its problematic meaning, they can significantly influence DO concentrations in coastal regions and especially on continental shelves at eastern oceanic boundaries where summer-time (Northern Hemisphere) upwelling advects low-oxygen waters from deep to near-surface layers.

The hypoxic conditions are considered those when a marine ecosystem experiences stress from the lack of oxygen and sea-living organisms start to die. The hypoxia criterion may vary as it depends on what species are regarded. For example, some benthic worms can tolerate DO concentrations as low as 0.7 ml/l and for the striped bass DO=3 ml/l could be already lethal [Scientific Assessment of Hypoxia, 2010]. Diaz and Rosenberg [1995] define an average DO value for hypoxia threshold as DO=1.43 ml/l. In present study we use DO=1.43 ml/l as a hypoxia and DO=0.5 ml/l as an anoxia thresholds following the majority of investigators (e.g. [Connolly et al., 2010], [Chan et al., 2008], [Grantham et al., 2004]). Under certain physical and biological conditions, hypoxic waters may occupy significant areas of the continental shelf (defined here as the area bordered offshore by the 200 m isobath) and impose a big threat to sea-living organisms especially immobile or relatively slow-moving species, such as mussels, clams, and crabs that cannot escape from a deoxygenated zone. If the onset of hypoxia waters is too rapid, some fishes may be trapped in low-oxygen area and also die.

Coastal ocean hypoxia may result from several different processes. The most important source is the anthropogenic eutrophication that results from runoff of nutrient-rich agricultural and wastewaters into coastal areas. Strong density stratification that prevents vertical mixing and oxygen exchange between atmosphere and underlying dense waters is the cause of natural hypoxia. The world's largest area of bottom hypoxia (ca. 50,000 km2) in the Baltic Sea is a result of strong density stratification [Savchuk, 2010]. The second largest coastal ocean hypoxia area is the continental shelf of Texas/Louisiana, where runoff from the Atchafalaya and Mississippi rivers overload a shallow shelf that has relatively weak circulation with nutrients washed off from agricultural fields [Hetland & DiMarco, 2008]. Hypoxia on the shelves adjacent to eastern boundary currents (Humboldt, Benguela, California) is generally associated with upwelling circulation that transports nutrient-rich and oxygen-poor deep waters onto the shelf. In this case, hypoxia could emerge as a result of two mechanisms: (i) low-oxygen upwelled water dilutes resident waters and (ii) upwelled nutrient-rich water enhances biological production, much of which eventually sinks to the bottom as detritus and consumes oxygen as it decomposes.

While hypoxic and anoxic events have occurred regularly in the Humboldt (off Peru) and Benguela (off Namibia) Currents, it was only in the last decade that hypoxic and anoxic events have been reported on the continental shelf off Oregon coast. Chan et al. [2008] provide an overview of historic records of DO on the Oregon shelf from 1950 – 2006. According to their data, there had not been significant hypoxia on the Oregon shelf prior to 2000. Subsequently, several strong hypoxic events occurred during the interval from 2000 to 2006. In particular, inner-shelf anoxia and shelf-wide hypoxia occurred during summer-fall of 2006. Grantham et al. [2004], based on surveying across Central Oregon shelf, report severe inner-shelf hypoxia during summer 2002 with registered DO concentrations as low as 0.21 ml/l.

Hypoxia is a result of interactions between physical and biological processes. The complexity of these interactions does not allow the exact mechanisms responsible for hypoxia to be determined from observations only. Spatial and temporal coverage of the observations in the early 2000s (e.g. used in Chan et al. [2008], Grantham et al. [2004], and the present study GLOBEC-LTOP) does not permit assessment of the importance of physical versus biological processes and the impact of remote forcing such as from offshore and northern regions during the upwelling season. State-of-the-art biological-physical modeling could be a better approach for characterizing the roles of biological and physical processes in creating shelf hypoxia, as suggested by Pena et al. (2010) previously. The ultimate goal of this study is to determine the causes of hypoxia on the Oregon shelf using a coupled physical-ecosystem model.

**Methods: Coupled Ecosystem-Physical Model**

In order to achieve this goal we develop a coupled biological-physical model for the coastal ocean off Oregon. We coupled the Regional Ocean Modeling System (ROMS v3.0) physical model [Shchepetkin and McWilliams, 2003, 2005] with a 5-component NAPZD (Nitrate, Ammonium, Phytoplankton, Zooplankton, Detritus) ecosystem model, based on Spitz et al. [2005]. The ROMS Coastal Transition Zone (ROMS-CTZ) computational domain spans over (129-124 W) in meridional and over (40.5-47.5

N) in zonal directions, and is identical to that used in Koch et al. [2010]. The horizontal grid has roughly 3 km resolution, and the 40 sigma-levels in the vertical are arrayed to produce finer resolution in the surface and bottom boundary layers. Bottom topography is composed by merging two sets: a high-resolution (12″) NOAA-National Geophysical Data Center (NGDC) bathymetry of the U.S. West Coast, representing features on the shelf and continental slope, and a lower-resolution (5′) ETOPO5 product [NGDC, 1988]. A minimum depth of 10 m is set along the coastline. Following Koch et al. [2010] we simulate the period from April through August, which includes the transition in spring from northward winds to southward upwelling-favorable winds and summer upwelling circulation.

The biological state variables are treated as tracers in exactly the same way as temperature and salinity, so are subject to advection and mixing, in addition to the biological processes.. We added dissolved oxygen (DO) to the Spitz et al. [2005] NAPZD model to create a 6-component NAPZDO model (Appendix A). DO is treated as a passive tracer with biologically mediated inputs (photosynthesis) and losses (zooplankton respiration, detritus remineralization and oxidation of ammonium), and an additional source (sink) term through air-sea exchange. The computation of air-sea DO flux uses DO saturation concentration after Garcia and Gordon [1992] and gas transfer coefficient after Keeling et al. [1998].

We generate solutions for three spring summers. The spring-summer of 2002 and 2006 had extensive and severe hypoxia on the central Oregon shelf for most of the summer, and 2008, when shelf hypoxia occurred late and was not widespread.

**Atmospheric forcing, initial and open boundary physical conditions**

The coupled model simulations for 2002 are forced by 9-km horizontal resolution daily-averaged COAMPS [Hodur, 1997] winds and monthly-averaged NCEP/NCAR [Kalnay et al., 1996] fields for heat-flux computation (short-wave solar radiation, air temperature, air pressure, relative humidity, precipitation) with 2.5 degrees horizontal resolution. The simulations of 2006 and 2008 were forced using 9-km atmospheric fields from the North American Mesoscale (NAM) model.

Initial conditions and open boundary conditions for velocities, temperature, salinity, and sea surface height (SSH) were provided by a larger scale NCOM-CCS model with 9-km horizontal resolution and 40 vertical levels: 20 sigma-levels (in upper 150 m) and 20 z-levels with constant depths [Shulman, 2004]. Open boundary fields are provided daily. The NCOM and ROMS domains are shown in Figure 1. To suppress undesired effects of open boundary conditions as a result of merging larger-scale 9-km horizontal resolution NCOM fields with smaller-scale 3-km ROMS fields we implemented a sponge layer that provided enhanced diffusivity and dissipation in the 100 km region adjacent to the open boundaries. Our analysis of biological and DO fields was restricted to the sub-domain from 43.5–46.5°N that excludes the sponge layer.

**Initial and open boundary ecosystem conditions.**

The open boundary conditions for nitrate, ammonia, phytoplankton, zooplankton and detritus are provided by the NCOM-CCS biological solutions for April-August of 2006 and 2008. NCOM simulations for 2002 did not include the ecosystem model, so for 2002 ROMS-CTZ we used NCOM-CCS biological fields from 2008, since wind forcing that year (more similar to 2002 than was 2006 winds), and 2008 had only mild hypoxia on the Oregon shelf. was moderate during 2002. NCOM's ecosystem has two types of phytoplankton (diatoms and nannoflagellates) and zooplankton (microzooplankton and mesozooplankton). Since the amounts of nannoflagellates and micro-zooplankton in NCOM fields are much less than the amounts of diatoms and mesozooplankton, respectively, we used the sum of the phytoplankton and zooplankton for our model.

The initial nitrate conditions for the NCOM-CCS ecosystem model are based on Levitus World Ocean climatology. During the several decade long spin-up of NCOM-CCS the nitrate fields showed significant drift, and by the 2000s both the nitrate concentration and depth of the nutricline had become biased (Igor Shulman, personal communication). To eliminate this bias in NO3 fields is crucial for realistic simulation of shelf ecosystem processes including oxygen dynamics. We eliminated the nitrate bias by adjusting the NCOM nitrate fields using empirical linear regression. The coefficients of linear regression were computed by comparing NCOM-CCS and GLOBEC-LTOP [Strub et al., 2002; Wetz et al., 2004] nitrate values taken at the same locations of the space-time domain for 8 depth layers (0-50, 50-100, 100-150, 150-250, 250-350, 350-500, 500-700, 700 m-bottom). The correction was done separately for NCOM-CCS data from 2006 (Figure 3a) and 2008 (Figure 3b). GLOBEC-LTOP data on NO3 and DO were collected at standard depths spanning from surface to 1000 m depth (or bottom if shallower) along traditional Oregon observation lines: Crescent River (CR), Rogue River (RR), Five Miles (FM), Heceta Head (HH), and Newport Hydrographic (NH, Figure 1) extending from the inner-shelf offshore as to 126°W. Since the NCOM model did did not include DO dynamics, we used a NO3:DO linear regression to estimate the DO field from NO3 for both boundary and initial conditions. The NO3:DO linear relationship was more robust than other relations between DO and density, temperature or salinity. The linear regression ratio was derived using all GLOBEC-LTOP NO3 and DO GLOBEC-LTOP observations from March-April of 1997-2004 (Figure 4). The GLOBEC-LTOP observational program ended in 2005, so 2006 and 2008 are missing. We assumed that the NO3:DO relation derived from 1997-2004, applied also to 2006 and 2008.

It is critical (as will be shown later by "Sensitivity analysis") that the simulation begin with realistic DO and NO3 concentrations in order to reproduce the shelf ecosystem dynamics including hypoxia correctly. This is why we took great care to provide initial NO3 and DO fields as close to reality as possible. Initial conditions for phytoplankton, zooplankton, and detritus were provided from NCOM-2006 fields (2006) or NCOM-2008 (for 2002 and 2008). Initial DO and NO3 for 2002 came from 2002 LTOP. For 2006 and 2008, NO3 came from the LTOP multiyear climatology. Initial DO for 2008 also came from the LTOP climatology, but the initial spring 2006 DO came from a shelf-wide survey (Jay Peterson, Oregon State University, unpublished data).

**Model-data comparisons.**

Detailed comparisons of model physical results with observations are described for 2002 in Koch et al. [2010]. There was good model-data agreement on the structure and seasonal development of surface and depth-averaged velocities over the shelf and in the offshore transition zone, the structure and development of the upwelling SST front, the separation and offshore intensification of the upwelling jet, and the 3-dimensional density field.

For the ecosystem components we compare the modeled vertical profiles of NO3, DO and Chl-a to GLOBEC-LTOP July 2002 vertical profiles (e.g., 3.5 months after the start of the simulation). Figure 5 shows vertical profiles of NO3 and DO along the NH line for 10-12 July 2002. There is a very good agreement between modeled and observed NO3 and DO at offshore deep locations and on the shelf. There is a region over the continental slope (NH25, NH35), where the NO3 and DO profiles from the model underestimate DO and overestimate NO3 at intermediate depths, although at both shallower and deeper depths the model-data agreement is quite good.

A vertical section of modeled and observed phytoplankton biomass (converted from chl-a using a conversion of 1.5 mmol N m$^{-3}$ [mg chl-a m$^{-3}$]$^{-1}$) from July 2002 show good agreement (Fig. 6a). Both show a surface maximum in the inner shelf and a subsurface maximum at about 20 m farther offshore. The comparisons of observed and modeled DO (Figure 6b) show both qualitative and quantitative agreement with two spatial maxima: near-surface onshore associated with upwelling-induced surface coastal phytoplankton maximum and around 20 m depth offshore (around 124.7°W), associated with a subsurface phytoplankton maximum seen on Figure 6a.

**Results.**

**Description of summer shelf hypoxia in 2002, 2006, and 2008.**

We start the analysis of summer shelf hypoxia off Oregon with the scenarios which have the most realistic initial and boundary conditions for ecosystem components. In Figure 7 we show modeled summer hypoxia development on the Oregon shelf in (a) 2002, (b) 2006, and (c) 2008. The top panels of Figure 7 show time series of shelf water volume (or is this the percent of the shelf bottom that is hypoxic) with hypoxic DO concentrations. The bottom panels of Figure 7 show percentage of cross–shelf area with hypoxic conditions as a function of time and longitude. The characteristics of summer shelf hypoxia for the three years – means and minima of DO concentrations in hypoxic waters both at the bottom and over the entire vertical extent of hypoxia, means and maxima of both bottom area and volume of hypoxic waters – are presented in Table 2.

For 2002, shelf hypoxia first appeared in the model during mid-June near 44°N (south Heceta Head (HH) complex – shelf area between 43.9 – 44.5°N) from where it spreaded north, steadily overtaking the entire shelf north of 44°N; hypoxic conditions occurred in up to 28% of total shelf volume (Figure 7a). Of the three years modeled, summer hypoxia in 2002 is the longest, lasting for 105 days. Hypoxic waters were widespread in 2002, and occurred on up to 67% of the bottom shelf area. The mean near-bottom DO concentration is 1.33 ml/l and the lowest DO concentrations were 0.07 ml/l (Table 2). In 2006, hypoxic waters first occur in the northern part of the shelf between 45.5 – 46°N (Fig. 7b), but <10% of the shelf is hypoxic. Hypoxia of >50% of the shelf at a specific latitude doesn't happen until late July near the Heceta Head line (south of 44.7°N). The model suggests that hypoxia in 2006 occurred later, but was more prevalent across the shelf once it started than in 2002. Hypoxia in 2006 lasted for at least 82 days (still present at the end of simulation). The mean and minimum simulated DO concentrations in 2006 are 1.35 and 0.27 ml/l, respectively (Table 2). In 2008, modeled hypoxia on the Oregon shelf had the shortest duration (55 days) and the least extent both at the bottom (up to 40% of shelf area) and over the entire water body (up to 26% of total shelf volume, Table 2). Hypoxia first appears near Heceta Bank (ca. 44.5°N) during late July, and rapidly is found at all latitudes to 46N by early August. However, over most of the shelf, hypoxia lasted only until the middle of August (Figure 7c). Curiously, the lowest DO concentrations in all three years were observed not at the bottom but in upper layers.

The analysis of Figure 7 and Table 2 show that the characteristics of summer hypoxia – timing, spatial extent, severity, and geographical distribution – are very different for the three years. To better understand these interannual differences in hypoxia development we conducted additional numerical experiments.

**Sensitivity analysis experiments.**

Since hypoxia is more pronounced in 2002 and 2006 than in 2008, and we have more in situ data for those years, we examined the effects of variable NO3 and DO initial and boundary conditions on the hypoxia development. The details of initial and open boundary conditions for ecosystem components and ocean physics and atmospheric forcing for principal numerical experiments are given in Table 1. The base case scenarios, BC2 (2002; BC for Base Case) and BC6 (2006), described above used the most realistic initial and open boundary conditions. For initial conditions these were LTOP-2002 for NO3 and DO (2002) and Peterson in situ observations and LTOP-Clim for DO and NO3, respectively in 2006. Experiments UI2 (2002; UI for Unmodified Initials) and UI6 (2006) use the BC boundary values for NO3 and DO but different initial NO3 and DO conditions based on the unmodified NO3 NCOM fields. Another set of simulations, CI2 (2002; CI for Climatological Initials) and CI6 (2006), use the same boundary conditions but initial conditions formed from the 1997-2004 GLOBEC-LTOP climatology. The final set of experiments, UB2 (2002; UB for Unmodified Boundary) and UB6 (2006) use the same initial conditions as BC2 and BC6, respectively, but boundary NO3 conditions from unmodified NCOM fields. The 2008 summer, because it was the year with weakest hypoxia development and fewest in situ observations for evaluation of the model, was not included in the sensitivity study, and we show only the Base Case (BC8) that used our best estimate of initial and boundary DO and NO3 conditions.

**Analysis of the basic simulations**

In order to assess summer hypoxia development on the Oregon shelf and compare its characteristics among numerical experiments with different conditions, for June, July and August, we determine the number of days per month when each bottom location in the model was hypoxic (Fig. 8). First, we analyze hypoxia development for experiments BC2 (2002) and BC6 (2006), Figure 8c, g, respectively, when we have the most realistic model set-up in terms of initial and boundary DO and NO3 conditions.

As previously noted from Figure 7c, hypoxic waters in 2002 first occur on Heceta Bank in June. Heceta Bank is a relatively shallow shelf area encircled by deeper bathymetry. The coastal jet veers offshore near Heceta Bank. Flows above the bank are generally sluggish (REF), and provide favorable conditions for phytoplankton production to accumulate and be consumed by zooplankton or sink as detritus to depth. Near the bottom, DO is reduced through zooplankton mortality, detritus decomposition, and ammonium oxidation. Because of the long retention time on the bank, bottom waters are more likely to reach hypoxic DO concentrations. We examined the magnitude of oxygen equation processes to assess their importance to hypoxia development. Figure 9 shows the rate of DO change due to physical (advection and diffusion) and biological (zooplankton mortality, ammonium oxidation, and detritus decomposition) processes, their combination (the net rate of DO change), and the DO concentration of bottommost layers for each summer month of 2002 and 2006. On the outer part of Heceta Bank (ca. 44N), biological consumption of DO dominates over slight DO increase due to physics, to produce an overall decline in DO (Figure 9Ac). In July, hypoxic conditions expanded northward along the shelf break and hypoxic waters occupy the whole of Heceta Bank, from the shelf break to the coast. Coastal waters are affected by hypoxia up to almost 45 N (Figure 8c). In August, bottom hypoxia occurs along the entire shelf from 44-46N. The distribution of the net DO forcing at the bottom in July and August (Figure 9Bc,Cc) is consistent with the longer retention times derived from Figure 8c. It shows declining DO forcing over the shelf mostly due to biological forcing, esp. decomposition of organic matter at depth (Figure 9Bb,Cb) while physical forcing tends to be positive (e.g., increasing DO; Figure 9Ba,Ca). Physical processes are enhancing DO production in the south sufficiently such that by August, offshore regions of Heceta Bank are no longer hypoxic.

In 2006, bottom hypoxic waters first appear in June in the northern part of the Oregon shelf along the shelf break between 45.5 – 46 N (Figure 8g, left panel) suggesting advection of low oxygen deep waters from offshore and the importance of northern DO and NO3 boundary conditions. In July, the hypoxic conditions expanded southward to 44.7 (Figure 8g, middle panel) occupying bottom waters adjacent to the shelf boundary. In July, there is another region of hypoxia nearshore between 44.3 – 44.9 N, which is dissociated from the low DO waters advected from offshore. The dissociation of the two low DO pools suggests that the nearshore may be a consequence of an onshore phytoplankton bloom from upwelling circulation. This is supported by the biological consumption of oxygen around 44.5 N that grows through summer 2006 (Figure 9Db,Eb,Fb). In August, the two pools of low DO bottom waters converge and occupy the most of the shelf from 44.3°-46°N and from the shelf break to the coastline (Figure 8g, right panel). The general pattern of physical and biological DO drivers development at the bottom is similar to the 2002 case. Biological forcing is negative and increases with time while physical forcing, negative in June, becomes positive (bringing high DO water to Heceta Bank) and increases in the south in July and August preventing hypoxia from spreading over the Bank and farther south (Figure 9D-F).

Judging from the relative rates and signs of physical and biological DO changes at the bottom, the more important determinant of hypoxia is biological forcing, i.e. detritus decomposition. However, if we consider instead the whole water column, it is not necessarily the case. We integrate the DO forcings due to physical and biological processes for the whole simulation interval and for the summer of 2002 and 2006 over the shelf (Table 3). Although DO reduction due to biological sink terms important at deeper layers, is comparable to the net DO loss in both years, the net biological term is positive and large owing to a very high rate of DO production through carbon fixation (photosynthesis) by phytoplankton in the photic zone (0.8733 and 0.8744 ml $O_2*10^{16}$ in 2002 and 2006, respectively, Table 3). Indeed, the physical mechanism of DO reduction, especially through horizontal advection and diffusion, resulting in 0.6215 and 0.9737 ml $O_2*10^{16}$ loss in 2002 and 2006, respectively (Table 3), is the critical factor for the negative balance of DO in the shelf area. The net loss of DO in shelf waters in the model is estimated as 0.4707 ml $O_2*10^{16}$ in 2006 which is 2.5 times higher than in 2002 (0.1785 ml $O_2*10^{16}$). Since the DO change due to biological processes was similar in 2002 and 2006, the large difference between years in net DO change is ultimately due to the difference in physical DO forcing. In 2002, the DO loss was distributed nearly evenly in time, whereas in 2006 about two thirds of total DO loss occurred in April and May (Table 3). Figure 10, where time series of the DO fluxes (physical, biological, and net) integrated over the shelf are shown, clearly represents this behavior as well as the close matching of physical and net DO fluxes in both years. In Figure 10, along with DO fluxes, we plot upwelling index representing coastal upwelling intensity along the coast in volume flux units. There is a strong negative correlation between DO physical flux and upwelling index in both years: -0.77 in 2002 and -0.68 in 2006. This dependency could be explained by a simple conceptual model. In an upwelling event, surface waters with relatively high DO concentration are advected offshore past the shelfbreak while deeper waters with lower DO concentration are advected onshore from the area outside the shelf. This circulation decreases DO concentration in shelf waters. On the other hand, deeper waters with high nutrient concentrations advected on the shelf facilitate phytoplankton growth and DO production, but apparently the physical component dominates.

**Simulations using modified initial or boundary conditions.**

The importance of initial conditions in early spring and open boundary conditions of DO and NO3 for hypoxia occurrence in summer of 2002 and 2006 is seen by comparing numerical experiments having different initial and boundary conditions (Figure 8).

When the initial DO and NO3 conditions are altered from the most realistic values to conditions derived from unmodified NCOM fields, neither 2002 (UI2, Figure 8a) nor 2006 (UI6, Figure 8e) show development of hypoxia on the Oregon shelf, except for a very brief episode of hypoxia in a small northeastern region in Aug 2006 (Figure 8e, right panel). There could be at least two reasons why hypoxia does not develop under such conditions. First, advection of high (overestimated) DO offshore waters onto the shelf might buffer the decline in DO due to biological processes sufficiently that hypoxia thresholds are not reached. Second, deeper than normal nitrocline prevents nitrate–rich waters to upwell on the shelf and to provide more favorable conditions for the phytoplankton production and subsequent oxygen depletion. Here we should note that the ratio between primary and secondary production in our simulations roughly equals 3, this estimate is robust through simulation interval and different years, however in initial and boundary NCOM fields this ratio is roughly 1. When initial DO and NO3 conditions are substituted for fields formed from the LTOP climatology, the solution still produces shelf hypoxia, though the difference from basic cases for both years 2002 (CI2) and 2006 (CI6) is evident (Figure 8b,f). In 2002, hypoxia emerges along shelf break north of 45 N and, again, in the HH complex only in July thus being delayed for about a month (Figure 8b, middle panel). In August, distribution of bottom hypoxic waters over the shelf is similar to the basic case BC2, though hypoxia is not observed along shelf break south of 45 N (Figure 8b, right panel). In 2006, the bottom hypoxia timing and spatial pattern is closer to the basic case BC6 than in 2002 (Figure 8f) since initial DO, NO3 conditions in 2006 are less different from LTOP-climatology than in 2002. The hypoxia starts again in June in the north-western part of the domain and propagates south along shelf break in July when the coast-adjusted hypoxia pool emerges around 44.5 N (Figure 8f, middle panel). The hypoxia along shelf break in CI6 is far less intense than in BC6. In August, the spatial coverage of bottom hypoxic waters is similar to BC6 but the residence time is smaller (Figure 8f, right panel). Having analyzed the experiments CI2 against BC2 and CI6 against BC6, we conclude that the climatological early spring DO and NO3 initial conditions could be a good substitute for the years when data are missing (e.g., in 2008), although hypoxia timing could be shifted and hypoxia spatial extent could be underestimated.

**Importance of boundary conditions.**

The next set of experiments shows how DO and NO3 conditions at the open boundaries influence summer hypoxia on the shelf. The shelf hypoxia details for numerical experiments using unmodified NCOM fields at the open boundaries for 2002 and 2006 (UB2 and UB6, respectively) are shown in Figure 8d,h. In 2002, bottom hypoxia appears first along the shelf break and on Heceta Bank in June (Figure 8d, left panel), but later than in BC2. The development and propagation of bottom hypoxia in July and August resembles that of BC2 but the spatial extent and residence time of bottom hypoxic waters is considerably less (Figure 8d, middle and right panels). Even with unrealistically high DO and low NO3 at the open boundaries, UB2 simulated the beginning of hypoxia better than did CI2 which had more realistic boundary conditions, but used climatological initial conditions. That suggests the importance of having accurate initial conditions for DO and NO3 in early spring of 2002 to replicate the progression of observed summer hypoxia.

In 2006, the unrealistically high DO and low NO3 boundary conditions from NCOM prevent or substantially delay the development of shelf bottom hypoxia (Figure 8h). Hypoxic waters first appear over a very small inner shelf area near 44.5°N in late July, with more widespread inner shelf hypoxia in August (Figure 8h, middle and right panels). Outer shelf hypoxia associated with advection of low-DO water from offshore is limited and appears late compared to BC6 (Fig. 8g). Comparison of UB2 and UB6 also suggests that the importance of outer shelf sources of hypoxia varied between years (Fig. 8d,h). Overall, this comparison suggests that using realistic DO and NO3 boundary conditions is important for reproducing summer hypoxia in 2006.

In our model domain, boundary effects might result from conditions on the open northern, southern or western boundaries. A simulation (not shown) where DO and NO3 concentrations at the open western boundary were set to zero showed no difference in shelf bottom hypoxia to the BC cases. The western boundary 400 km from the coast is too remote to influence DO and NO3 concentrations on the shelf, given the five month simulation period, and the cross-shelf velocities of flow. Thus, the effects of DO and NO3 entering the shelf subdomain through perimeter (northern, southern or western boundaries) ultimately came from either the northern or southern open boundaries. This is not surprising given the much greater magnitude of alongshore flows than cross-shelf flows in the CCS.

To estimate the contributions of the open shelf boundaries (northern, southern and shelf-break) to the DO budget on the shelf only, we computed and integrated DO fluxes normal to these boundaries over April-August and June-August of 2002 and 2006 (Table 4). Along with DO fluxes, we computed NO3 and volume fluxes. DO and NO3 fluxes are represented by advective and diffusive components. The volume flux is balanced so the net flux is zero, and it gives a good sense of the distribution of inflows and outflows among the boundaries. Volume fluxes through shelf boundaries were similar in both years. The most powerful inflow at the northern boundary is balanced by outflows at western and southern boundaries, where the former is approximately 5 times greater than the latter. The along-coast current was more energetic in 2002 with inflow at the northern boundary of ca. 4623 km$^3$ being 30% greater than in 2006 (ca. 3500 km$^3$).

The NO3 fluxes through northern (positive) and western (negative) boundaries in 2002 are approximately 50% and 100% larger than in 2006 (Table 4). This means, in 2002 waters with higher NO3 concentrations than in 2006 are advected to the shelf through northern boundary, but still higher NO3 concentrations waters are advected from the shelf offshore. This is explained by the fact that in 2002 background NO3 concentrations exceed that of 2006 as reflected in initial and boundary conditions. Higher rate of NO3 removal in 2002 together with larger NO3 inflow at southern boundary in 2006 makes net NO3 flux in 2006 somewhat larger than in 2002, but apparently nitrate is in excess and does not limit the biological DO production since it is almost exactly the same in both years (Table 3).

Dissolved oxygen fluxes, which are of the most interest here, are distributed at open boundaries in a similar proportion in 2002 and 2006. The ratio of 2002 and 2006 DO inflow through the northern shelf boundary is slightly higher than the ratio of volume fluxes. This supports the fact that DO concentrations at open boundaries are slightly higher in 2006. DO outgoing fluxes through western and southern boundaries normalized by volume fluxes in 2006 yet larger than in 2002 result in net DO flux being almost twice as large in 2006 (-0.5043 vs. -0.2591 ml $O_2*10^{16}$, Table 4). The reason lying behind that big difference in net DO fluxes in 2002 and 2006 is ultimately in higher upwelling rate in 2006 that provides more favorable conditions for physical DO removal from the shelf. Although winds were more energetic in 2002, the spring transition to upwelling-favorable winds in 2006 started earlier (first significant long-lasting upwelling event began in early April) and wind events lasted longer than in 2002 providing better conditions for more energetic upwelling (Figure 10). Upwelling index integrated over April-August is 1630 (2012) $km^3$ for 2002 (2006). It is worth to state here that twice as large DO loss in April-May, 2006 comparatively to June-August, already noted from Table 3, owes to the long and energetic upwelling event in April-May (Figure 10b).

The diffusive fluxes of both NO3 and DO are characterized by very low values at northern and southern boundaries and by values comparable to advective components at western boundary. This difference is explained by the fact that all ocean properties are much more homogenous in along-flow direction, and corresponding property's gradients are much lower than at the western boundary in cross-flow direction. Moreover, the horizontal mixing of tracers (NO3 and DO among them) is performed in our model along isopycnic surfaces that are nearly parallel to levels of constant depth in along-flow direction and inclined to them in cross-flow direction (where density gradient is significant) what makes cross-flow gradients of NO3 and DO even larger considering their strong depth dependency. Considering that diffusive fluxes at northern and southern boundaries are negligible compared to the western boundary flux, there is nothing to compensate it. This makes diffusive flux at western boundary yet more comparable to advective flux and more appreciable in its contribution to net flux (Tables 3, 4).

We have shown strong dependency of DO reduction in Oregon shelf waters on upwelling intensity. The low DO concentration waters flow onshore along the sloping bottom replacing more oxygenated water that is advected offshore. This less oxygenated water comes from the western shelf-break where it, in turn, was advected southward from the open northern boundary of the computational domain. Mean flow on the shelf and offshore in the Oregon CTZ in spring-summer is southward, which has been very well documented previously in models [e.g. Springer et al., 2009; Koch et al., 2010; this study] and observations [provide appropriate references; might start with Moorings, CODAR and esp. shipboard ADCP (Kosro refs)]. The physical mechanism behind the reduction of DO and bottom hypoxia development on the Oregon shelf in spring-summer is the intensive coastal upwelling. But the pattern and timing of hypoxia are very sensitive to having realistic DO and NO3 concentrations at the northern boundary (2006) and initial concentrations (2002), as shown by simulations (Fig. 8). In 2002, because of the anomalously low initial DO conditions observed by [see GRL special issue papers], on the other hand, even overestimated DO at the open boundaries did not prevent significant summer hypoxia from occurring (Fig. 8d).

In future studies it will be beneficial to track hypoxic water sources, possibly using Lagrangian model frameworks, to distinguish between waters advected directly by zonal flow and by upwelling circulation.

**Conclusions.**

A coupled biological-physical model based on a NPZD-type biological model and ROMS ocean model simulated dissolved oxygen (DO) dynamics and hypoxia development on the Oregon shelf during April-August of 2002, 2006, and 2008. We found that shelf hypoxia occurred during summer months of all three years. It was characterized by variable severity, horizontal and vertical extent, duration, and timing, and it was more pronounced in 2002 and 2006 than in 2008.

In order to identify the processes responsible for summer bottom hypoxia on the Oregon shelf in 2002 and 2006 we conducted a series of numerical simulations using variable initial and open boundary conditions of nitrate and dissolved oxygen. Based on the results of sensitivity analysis of these we conclude that (1) accurate forecasting of summer-autumn oxygen on the Oregon shelf requires having accurate ecosystem boundary conditions (especially for NO3 and DO) and late-spring initial conditions; (2) unrealistic initial DO and NO3 conditions in late-spring 2002 prevented or delayed hypoxia development; (3) realistic DO and NO3 conditions at the northern boundary are needed to accurately simulate summer hypoxia on the Oregon Shelf; this was especially critical for early bottom hypoxia on the shelf north of 45°N in 2006; (4) DO and NO3 conditions formed from "climatology" fields could serve as initial conditions when in situ data were lacking, but reliability of hypoxia predictions would be lessened, especially in years where initial concentrations diverged greatly from climatology (such as 2002); (5) modeled hypoxia occurred earlier in the north in 2006 and earlier in the south (Heceta Head) in 2002, perhaps, due to different northern boundary conditions for these years; (6) the DO and NO3 conditions at a western boundary located 400 km offshore did not have a significant impact on DO dynamics on the shelf in spring-summer.

Using the simulation model, we showed that DO changes due to biological processes (photosynthesis, respiration, remineralization) is large, although physical processes, mostly horizontal advection, is responsible for the net reduction in DO in spring-summer and the onset of bottom hypoxia in summer on the Oregon shelf. Coastal upwelling is the physical process most responsible for shelf hypoxia off Oregon. Diffusive fluxes of NO3 and DO are negligible at northern and southern boundaries of the Oregon shelf and appreciable at the western boundary. In 2006, about two thirds of total April-August DO loss happened in April-May as a result of strong and long-lasting upwelling favorable winds.

| Case\Factor | year | Initial Conditions | | | | Boundary Conditions | | | | Atm. forcing | |
|---|---|---|---|---|---|---|---|---|---|---|---|
| | | DO | NO3 | biology | physics | DO | NO3 | biology | physics | wind | heat flux |
| BC2 | 2002 | LTOP-2002 | LTOP-2002 | NCOM-2008 | NCOM-2002 | N:O on mod. NCOM'08 | NCOM:LTOP | NCOM-2008 | NCOM-2002 | COAMPS-2002 | NCEP-2002 |
| BC6 | 2006 | J.Peterson-2006 | LTOP-clim | NCOM-2006 | NCOM-2006 | N:O on mod. NCOM'06 | NCOM:LTOP | NCOM-2006 | NCOM-2006 | NAM-2006 | NAM-2006 |
| BC8 | 2008 | LTOP-clim | LTOP-clim | NCOM-2008 | NCOM-2008 | N:O on mod. NCOM'08 | NCOM:LTOP | NCOM-2008 | NCOM-2008 | NAM-2008 | NAM-2008 |
| UI2 | 2002 | N:O on NCOM'08 | NCOM-2008 | NCOM-2008 | NCOM-2002 | N:O on mod. NCOM'08 | NCOM:LTOP | NCOM-2008 | NCOM-2002 | COAMPS-2002 | NCEP-2002 |
| UI6 | 2006 | N:O on NCOM'06 | NCOM-2006 | NCOM-2006 | NCOM-2006 | N:O on mod. NCOM'06 | NCOM:LTOP | NCOM-2006 | NCOM-2006 | NAM-2006 | NAM-2006 |
| CI2 | 2002 | LTOP-clim | LTOP-clim | NCOM-2008 | NCOM-2002 | N:O on mod. NCOM'08 | NCOM:LTOP | NCOM-2008 | NCOM-2002 | COAMPS-2002 | NCEP-2002 |
| CI6 | 2006 | LTOP-clim | LTOP-clim | NCOM-2006 | NCOM-2006 | N:O on mod. NCOM'06 | NCOM:LTOP | NCOM-2006 | NCOM-2006 | NAM-2006 | NAM-2006 |
| UB2 | 2002 | LTOP-2002 | LTOP-2002 | NCOM-2008 | NCOM-2002 | N:O on NCOM'08 | NCOM-2008 | NCOM-2008 | NCOM-2002 | COAMPS-2002 | NCEP-2002 |
| UB6 | 2006 | J.Peterson-2006 | LTOP-clim | NCOM-2006 | NCOM-2006 | N:O on NCOM'06 | NCOM-2006 | NCOM-2006 | NCOM-2006 | NAM-2006 | NAM-2006 |

Table 1. Details of initial, open boundary conditions and atmospheric forcing.

| year | experiment | initial DO and NO conditions | open boundary DO and NO conditions | mean/min DO in hypoxic shelf waters (ml/l) | mean/max hypoxic shelf volume (km3) | mean/max hypoxic shelf volume (%) | mean/min DO in bottom hypoxic shelf waters (ml/l) | mean/max hypoxic shelf bottom area (km2) | mean/max hypoxic shelf bottom area (%) | hypoxia duration on shelf (days) |
|---|---|---|---|---|---|---|---|---|---|---|
| 2002 | BC2 | LTOP2002 | modif. NCOM'08 | 1.33/ 0.07 | 168.2/ 372.9 | 12.6/ 28.0 | 1.31/ 0.67 | 4621.9/ 8749.1 | 35.6/ 67.4 | 105 |
| 2006 | BC6 | J.P.<OP clim. | modif. NCOM'06 | 1.35/ 0.27 | 169.1/ 550.2 | 12.7/ 41.3 | 1.35/ 0.98 | 3397.5/ 7286.3 | 26.2/ 56.1 | 82 |
| 2008 | BC8 | LTOPclim | modif. NCOM'08 | 1.35/ 0.12 | 145.4/ 348.9 | 10.9/ 26.2 | 1.35/ 1.10 | 2209.1/ 5193.8 | 17.0/ 40.0 | 55 |

Table 2. Shelf hypoxia means and extrema.

| Factor/Time interval | **April-August 2002** | **April-August 2006** |
|---|---|---|
| Net | -0.1785 | -0.4707 |
| Physical=Advection+Diffusion | -0.5135 | -0.8637 |
| Air-Sea Flux | -0.1080 | -0.1100 |
| Physical+Air-Sea Flux | -0.6215 | -0.9737 |
| Biological Source | 0.8733 | 0.8744 |
| Biological Sink | -0.4303 | -0.3713 |
| Biological Net | 0.4430 | 0.5031 |
| Horizontal Advection | -0.3069 | -0.5739 |
| Horizontal Diffusion | -0.2061 | -0.2892 |
| Vertical Advection | -0.0006 | 0.0008 |
| Vertical Diffusion | -0.0001 | 0.0002 |
| Factor/Time interval | **June-August 2002** | **June-August 2006** |
| Net | -0.1190 | -0.1679 |
| Physical=Advection+Diffusion | -0.3339 | -0.4563 |
| Air-Sea Flux | -0.0829 | -0.0655 |
| Physical+Air-Sea Flux | -0.4168 | -0.5218 |
| Biological Source | 0.6061 | 0.6382 |
| Biological Sink | -0.3082 | -0.2844 |
| Biological Net | 0.2978 | 0.3539 |
| Horizontal Advection | -0.2328 | -0.3371 |
| Horizontal Diffusion | -0.1002 | -0.1183 |
| Vertical Advection | -0.0008 | -0.0011 |
| Vertical Diffusion | -0.0000 | 0.0001 |

Table 3. Time integrated DO shelf fluxes (ml $O_2$ *$10^{16}$).

| Time interval | April-August 2002 | | | | | April-August 2006 | | | | |
|---|---|---|---|---|---|---|---|---|---|---|
| | $NO_3$, mmol N*$10^{13}$ | | $O_2$, ml*$10^{16}$ | | Vol., $km^3$ | $NO_3$, mmol N*$10^{13}$ | | $O_2$, ml*$10^{16}$ | | Vol., $km^3$ |
| Boundary | advection | diffusion | advection | diffusion | | advection | diffusion | advection | diffusion | |
| Northern | 6.6094 | 0.0103 | 2.2439 | -0.0016 | 4623 | 4.1853 | 0.0101 | 1.8474 | -0.0016 | 3535 |
| Western | -5.0447 | 0.8499 | -1.9580 | -0.1285 | -3930 | -2.2775 | 0.9262 | -1.7371 | -0.1577 | -2903 |
| Southern | 0.2809 | -0.0059 | -0.5450 | 0.0013 | -693 | 0.7358 | -0.0081 | -0.6146 | -0.0015 | -632 |
| Net | 1.8456 | 0.8543 | -0.2591 | -0.1288 | 0 | 2.6436 | 0.9282 | -0.5043 | -0.1577 | 0 |
| Time interval | June-August 2002 | | | | | June-August 2006 | | | | |
| | $NO_3$, mmol N*$10^{13}$ | | $O_2$, ml*$10^{16}$ | | Vol., $km^3$ | $NO_3$, mmol N*$10^{13}$ | | $O_2$, ml*$10^{16}$ | | Vol., $km^3$ |
| Boundary | advection | diffusion | advection | diffusion | | advection | diffusion | advection | diffusion | |
| Northern | 4.1174 | 0.0062 | 1.3016 | -0.0009 | 2746 | 2.0651 | 0.0054 | 0.8977 | -0.0008 | 1728 |
| Western | -2.8458 | 0.4436 | -1.1510 | -0.0653 | -2308 | -2.9232 | 0.4334 | -1.3084 | -0.0683 | -2507 |
| Southern | 0.2659 | -0.0035 | -0.3603 | 0.0008 | -438 | 2.9074 | -0.0032 | 0.0974 | 0.0007 | 778 |
| Net | 1.5375 | 0.4464 | -0.2097 | -0.0654 | 0 | 2.0493 | 0.4357 | -0.3133 | -0.0685 | 0 |

Table 4. Time integrated $NO_3$, DO, volume fluxes through lateral shelf boundaries.

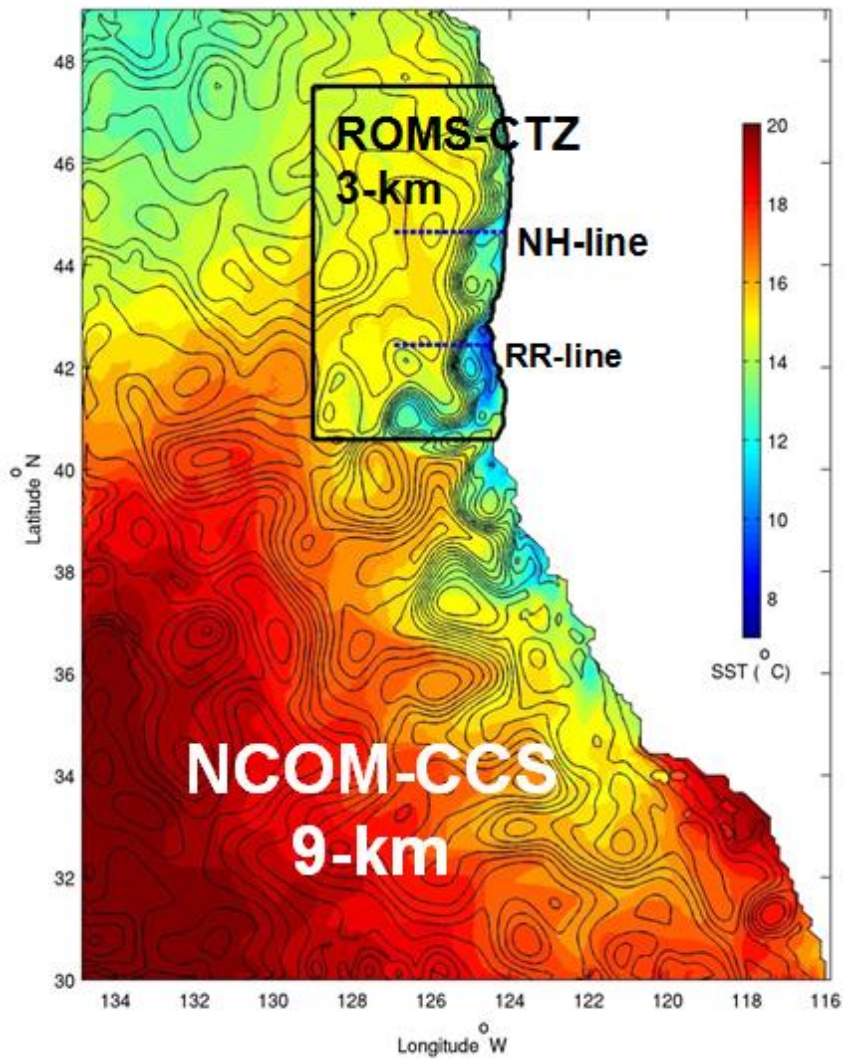

Figure 1. Computational domains of NCOM-CCS (9 km grid resolution) and ROMS-CTZ (3 km grid resolution) models. Fields of SST shown in color and SSH shown in contours represent good correspondence between outer and inner model solutions.

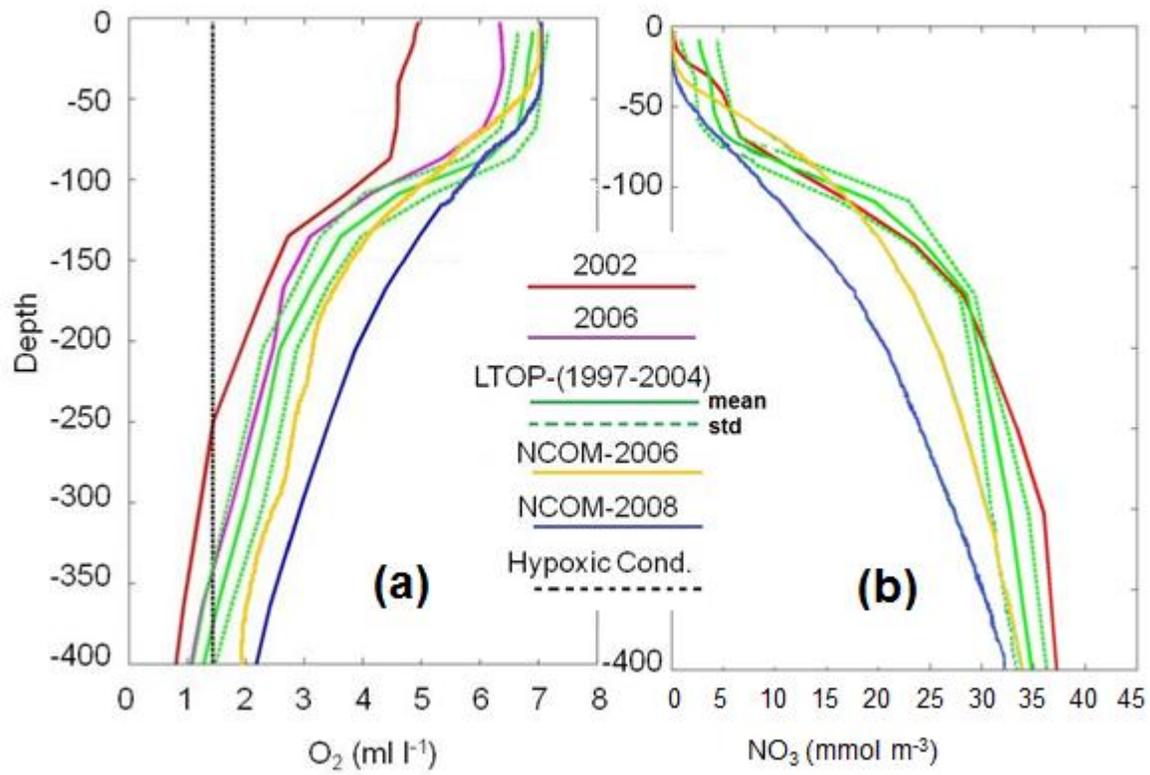

Figure 2. Early spring offshore (126 W, 44.6 N) profiles of DO (a) and NO3 (b) for different data sources.

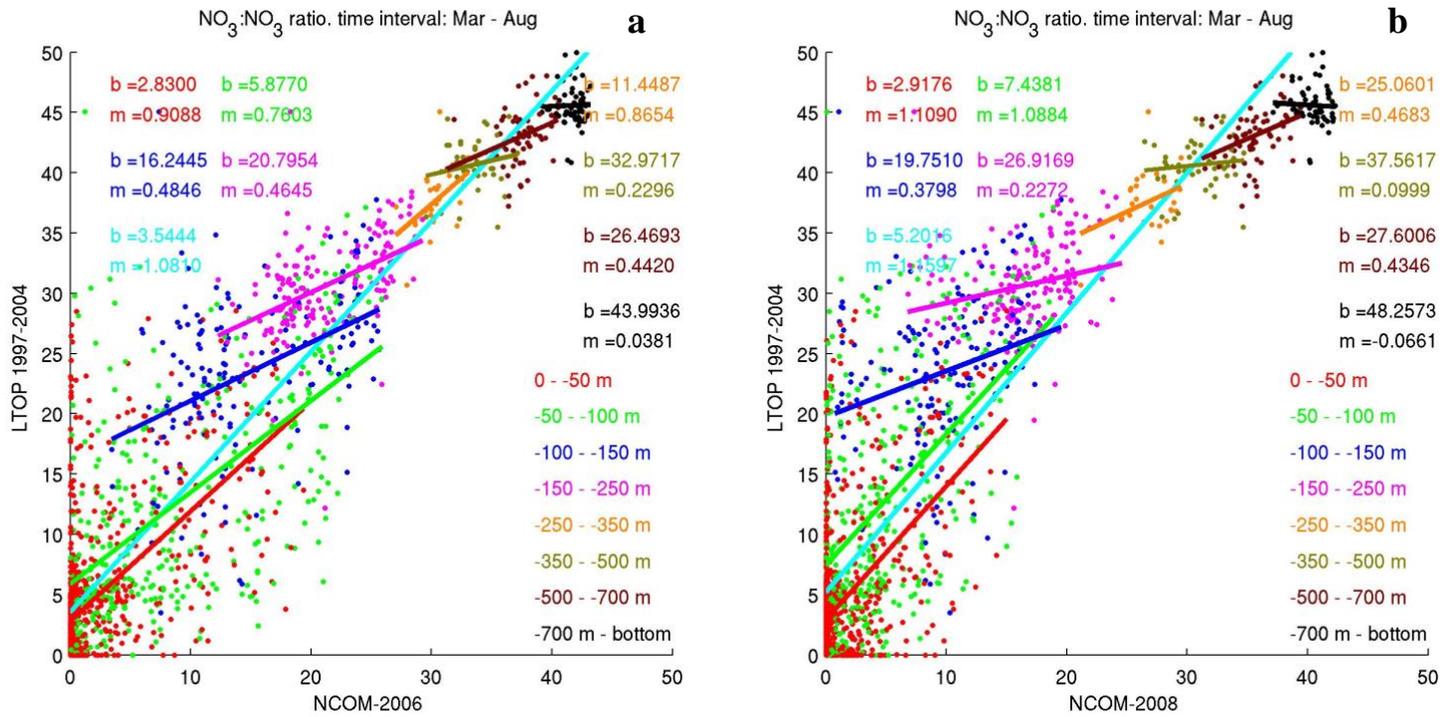

Figure 3. NO3:NO3 empirical ratios between GLOBEC-LTOP (1997-2004) and NCOM- (a) 2006, (b) 2008 data for March-August interval; m, b are linear regression coefficients.

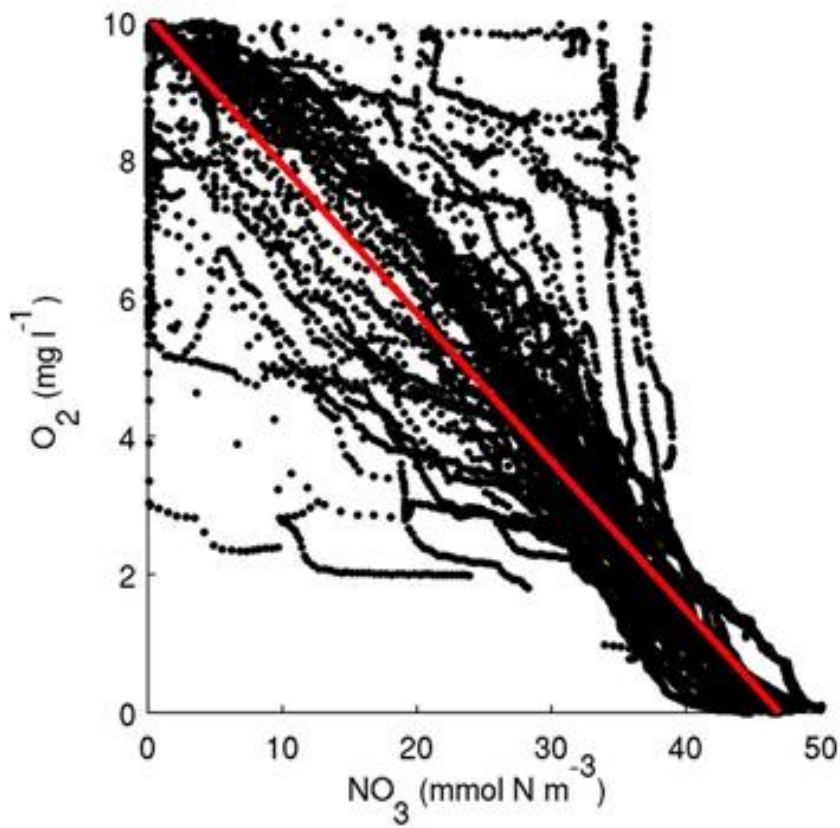

Figure 4. NO3:DO empirical ratio based on simultaneous GLOBEC-LTOP NO3, DO observations for March-August interval during 1997-2004.

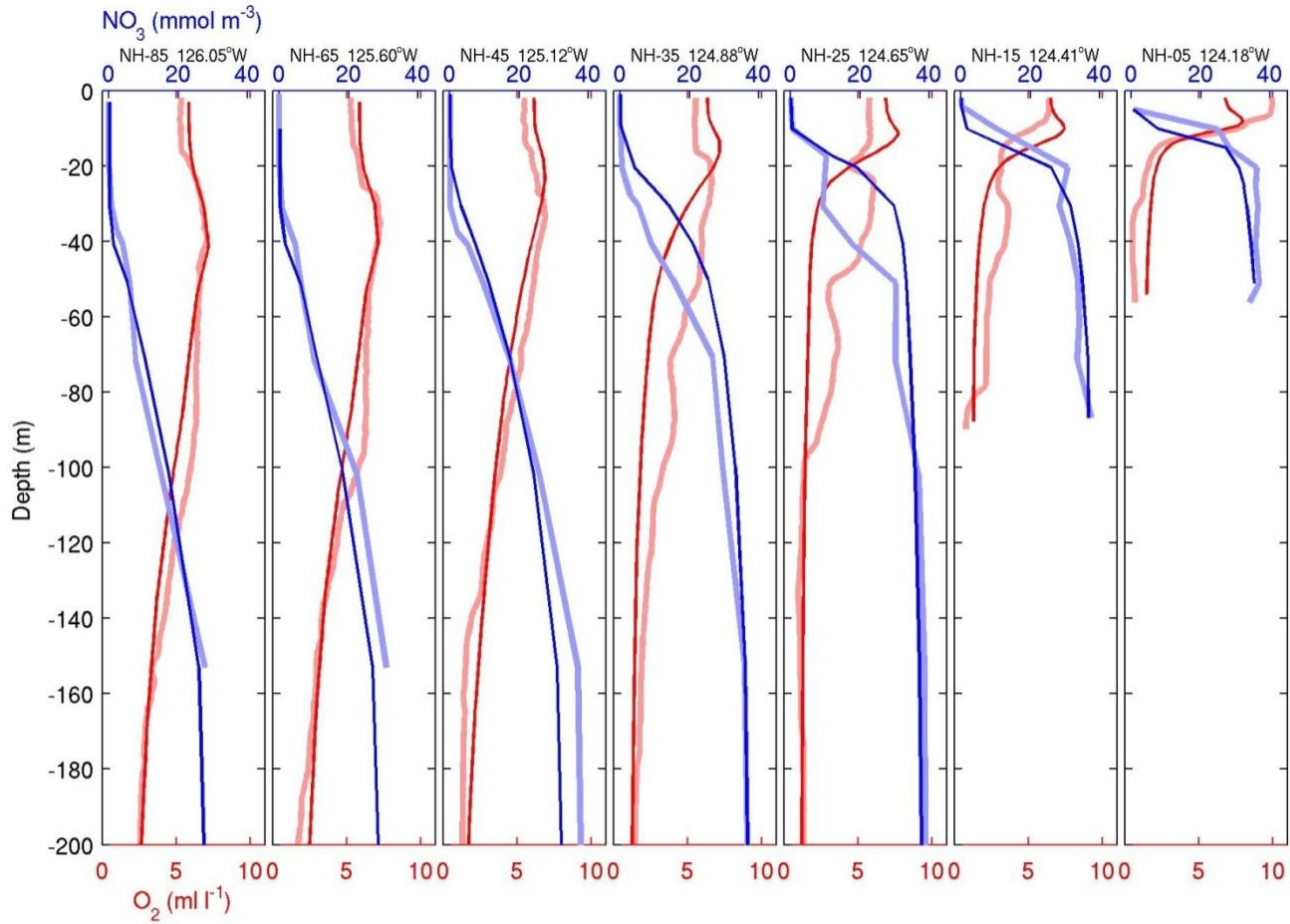

Figure 5. NO3 (blue) and DO (red) profiles along NH line (Figure 1, 44.65 °N) during 10-12 July 2002, NH-##: ##=offshore distance in miles, pale and thick lines represent data, bright and thin – model.

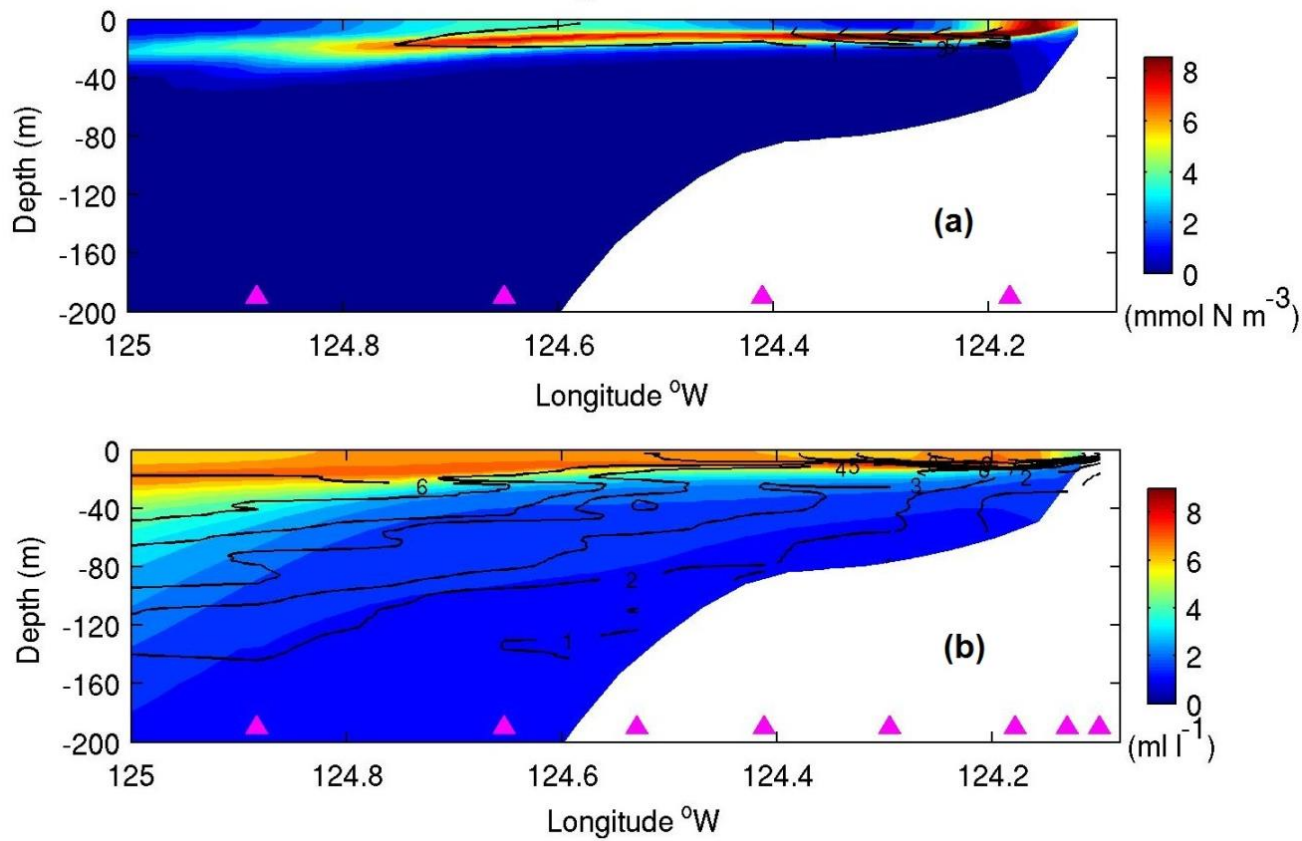

Figure 6. Across-shelf sections along the NH line (see Fig. 1) in July 2002 of (a) modeled phytoplankton (color) and observed chlorophyll-a concentrations (black contours) and (b) modeled (color) and observed (black contours) DO concentrations. Magenta triangles show the locations of LTOP stations where observations of chlorophyll and dissolved oxygen were done; note that chl-a profiles (extracted from Rosette Casts) were not done at all LTOP sites.

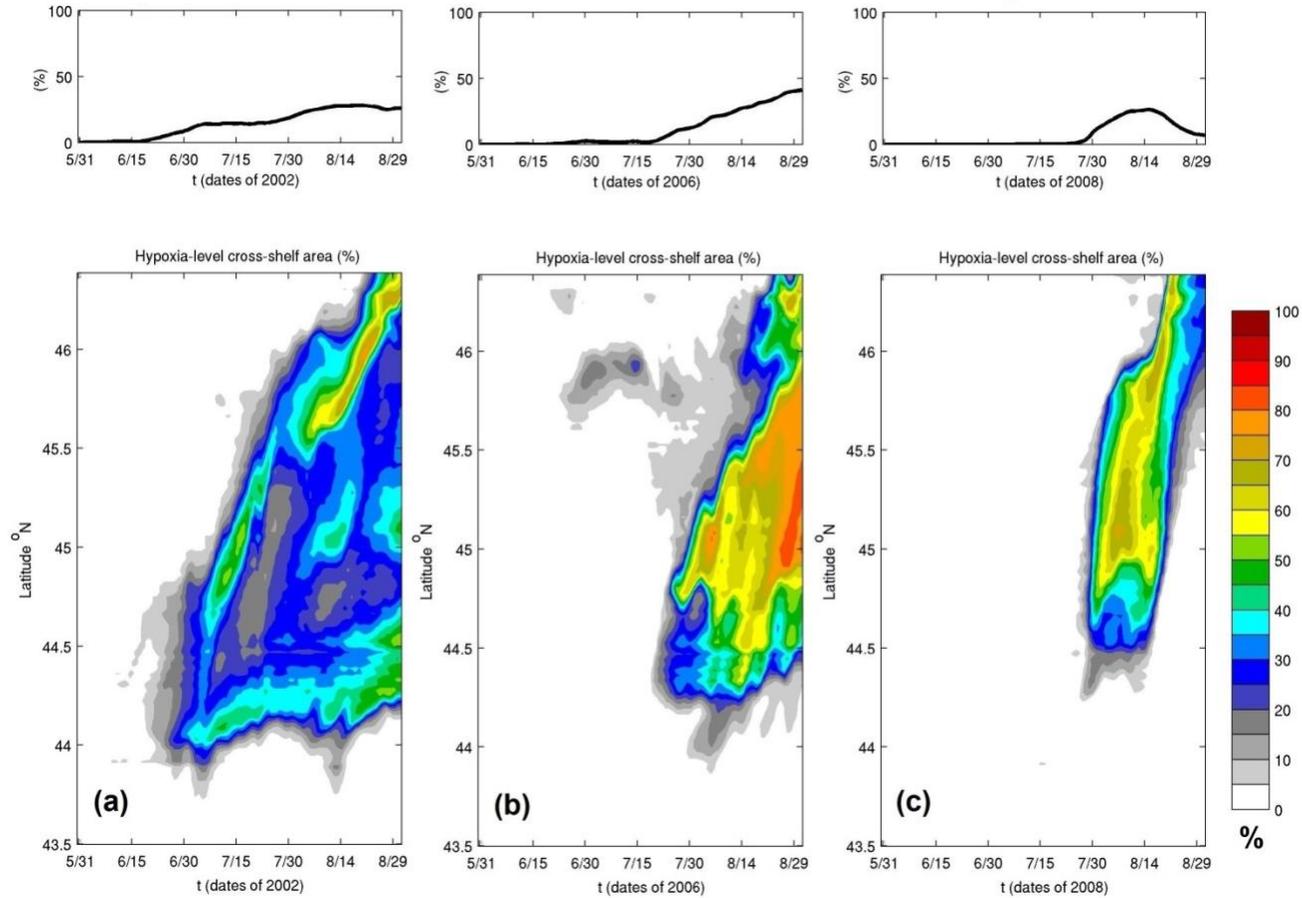

Figure 7. The time series of shelf water volume with hypoxic DO concentrations in % to total shelf volume (top panel) and the percentage of latitudinal cross–shelf area with hypoxic conditions as a function of time (bottom panel) for the base case model experiments (see Table 1): BC2 (a, 2002), BC6 (b, 2006), and BC8 (c, 2008).

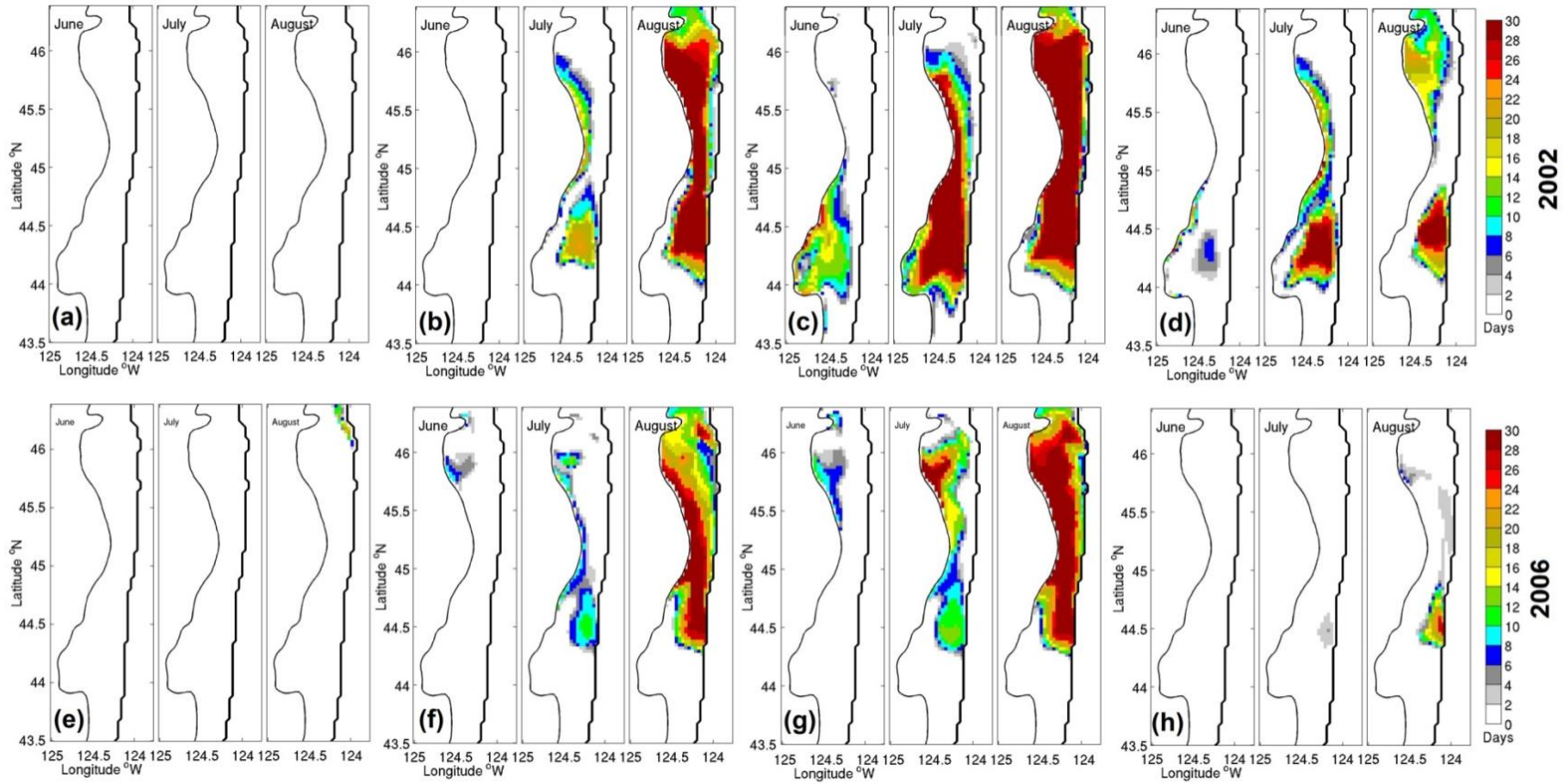

Figure 8. Number of days with hypoxic DO concentrations at the bottom over the shelf for the model experiments (a) UI2, (b) CI2, (c) BC2, (d) UB2, (e) UI6, (f) CI6, (g) BC6 and (h) UB6 (Table 1).

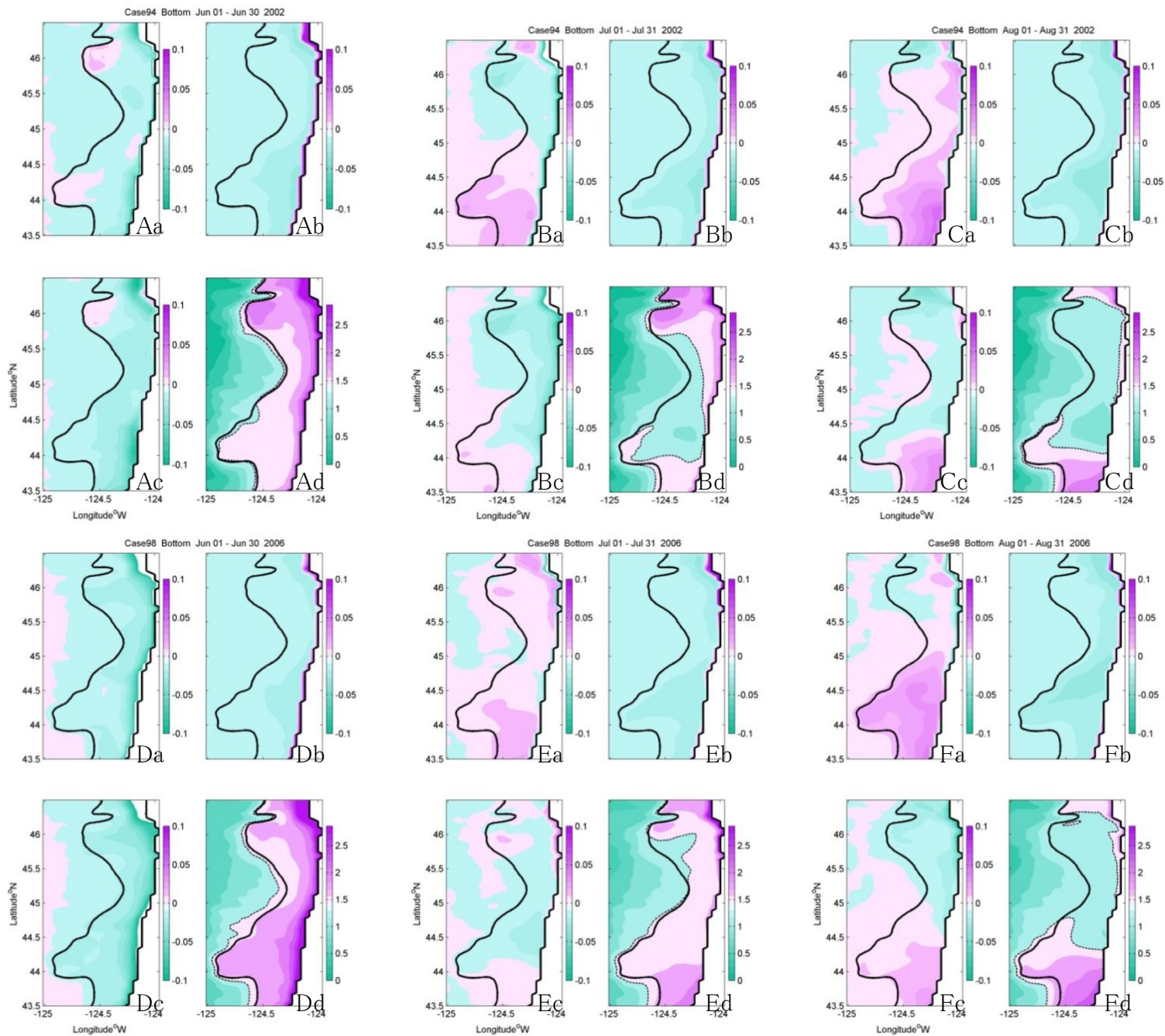

Figure 9. Rate of DO change due to (a) physical (advection and diffusion) and (b) biological (zooplankton mortality, ammonium oxidation and detritus decomposition) forcings, (c) their combination (the net rate of DO change) in ml l$^{-1}$ day$^{-1}$, and (d) DO concentration (ml l$^{-1}$) for bottommost layer averaged over (A,D) June, (B,E) July and (C,F) August of (A-C) 2002 and (D-F) 2006. Heavy black line is the shelf break at 200m depth.

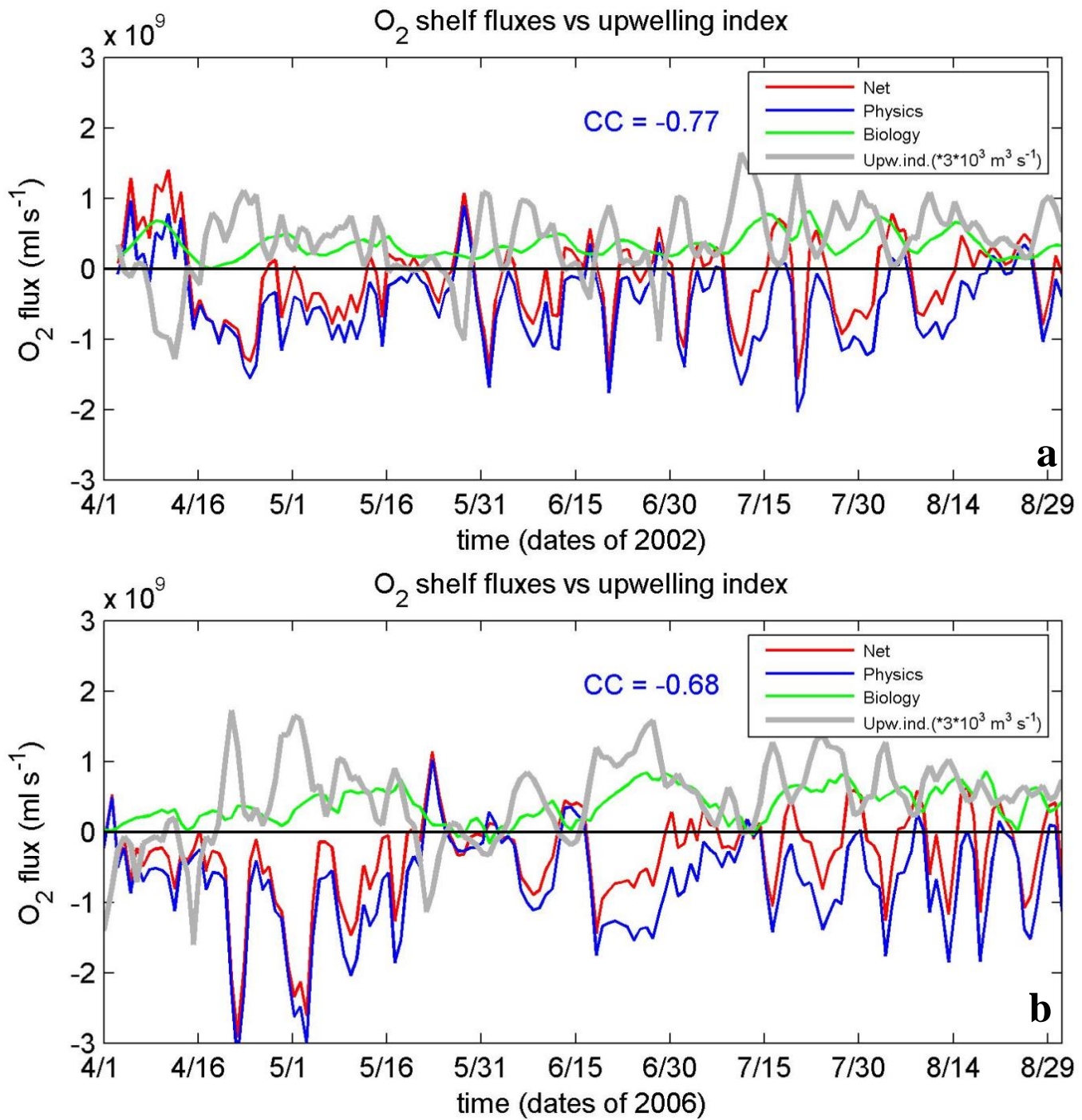

Figure 10. Physical (blue), biological (green), and net (red) fluxes of dissolved oxygen integrated over the shelf (ml s$^{-1}$) along with the zonal integrated upwelling index (NOAA, S. Pierce; *3000 m$^3$ s$^{-1}$) for (a) 2002 and (b) 2006. Upwelling index/physical flux correlations are shown.